\documentclass
[aps,pra,twocolumn,showpacs,floats,floatfix,amsmath]{revtex4}
\usepackage{graphicx}
\usepackage{amsmath}
\usepackage{amsfonts}
\usepackage{amssymb}%
\usepackage{epsf}
\usepackage{rotate}

\newcommand{\be}{\begin{equation}}
\newcommand{\ee}{  \end{equation}}
\newcommand{\ba}{\begin{eqnarray}}
\newcommand{\ea}{  \end{eqnarray}}
\newcommand{\ket}[1]{\left|#1\right>}
\newcommand{\bra}[1]{\left< #1 \right|}
\newcommand{\braket}[2]{\left< #1| #2 \right>}

\begin{document}

\title{Decoherence from Spin Environments} 
 
\author{F.M. Cucchietti$^{1}$, J.P. Paz$^{1,2}$ and W.H. Zurek$^1$}
\affiliation{(1): Theoretical Division, MS B213, Los Alamos National 
Laboratory, Los Alamos, NM 87545 \\
(2): Departmento de F\'\i sica, FCEyN, UBA, Pabell\'on 1, 
Ciudad Universitaria, 1428 Buenos Aires, Argentina}

\begin{abstract} 
We examine two exactly solvable models of decoherence -- a central 
spin-system, (i) with and (ii) without a self--Hamiltonian, interacting with 
a collection of environment spins. In the absence of a self--Hamiltonian
we show that in this model (introduced some time ago to illustrate 
environment--induced superselection) generic assumptions about the 
coupling strengths can lead to a universal (Gaussian) 
suppression of coherence between pointer states.
On the other hand, we show that when the dynamics of the central 
spin is dominant a different regime emerges, which is characterized
by a non--Gaussian decay and a dramatically different set of pointer
states.  We explore the regimes of validity of the Gaussian--decay
and discuss its relation to the spectral features of the environment 
and to the Loschmidt echo (or fidelity). 
\end{abstract} 


\maketitle 

\section{Introduction}

A central spin--system ${\cal S}$ interacting with an environment ${\cal E}$ 
formed by $N$ independent spins through the Hamiltonian
\be
{\cal H_{SE}} = \frac{1}{2}\sigma_z \otimes \sum_{k=1}^N g_k \sigma_z^{(k)}
\label{hamiltonian0}
\ee
may be the simplest solvable model of decoherence (we use the standard
notation according to which $\sigma_i^{(k)}$ and $\sigma_i$, $i=x,y,z$, denote 
Pauli operators acting on the $k$--th environmental spin and on the 
central system). This Hamiltonian 
was studied some time ago \cite{Zurek82} as a simple model of decoherence. 
It was used to show that relatively straightforward assumptions about the 
dynamics can lead to the emergence of a preferred set of pointer 
states due to environment--induced 
superselection (einselection) \cite{Zurek82,deco}.
Such models have gained additional importance in the past decade because of
their relevance to quantum information processing \cite{QIP}. 

The model described by (\ref{hamiltonian0}) was particularly useful 
to ilustrate the nature of decoherence in the context of a measurement. 
In such case the central spin is used as a simple (two state - one bit)
approximation for a memory of classical apparatus. Then, it is natural to neglect 
the effect of the system's self--Hamiltonian. As a consequence, the 
eigenstates of the interaction Hamiltonian (\ref{hamiltonian0}) emerge
as preferred pointer states of the system (defined
as the ones which are ``least perturbed'' by the interaction with the 
environment \cite{deco}). Thus, the eigenstates of the
$\sigma_z$ operator (denoted here as $|0\rangle$ and $|1\rangle$, with
eigenvalues $+1$ and $-1$ respectively) are 
dynamically selected by the interaction with the environment. 
Indeed, these states are not perturbed by the interaction while other
superpositions rapidly decay into their mixtures. 

Neglecting the self--Hamiltonian of the system is not always a reasonable
approximation. Studies of decoherence without such assumption have also
been carried out using, mostly, the Quantum Brownian Motion as a 
paradigmatic example \cite{QBM}. In such case, pointer states 
do not coincide with the 
eigenstates of the interaction Hamiltonian but can range from
coherent states for the QBM case \cite{sieve} to eigenstates of the 
system's Hamiltonian \cite{pz-1999}. Their properties are
determined by the interplay between the self--Hamiltonian and 
the interaction with the environment. 

In this paper we will study a generalization of the above simple model 
described by the Hamiltonian
\be
{\cal H_{T}} =\Delta \sigma_x + \frac{1}{2} \sigma_z \otimes 
\sum_{k=1}^N g_k \sigma_z^{(k)}.
\label{hamiltonian1}
\ee
This simple model includes both the effect of the evolution of the central
system and its coupling with the spin environment. 

The purpose of our study is twofold. First, in Section II we will analyze 
again the case where the central spin has no self--Hamiltonian 
($\Delta=0$ above). 
Our goal is to show that -- with a few additional natural and simple
assumptions about the distribution of coupling strengths $g_k$ -- 
one can evaluate the exact time dependence of the 
reduced density matrix of the central spin. In fact, we will 
demonstrate that the off--diagonal 
components display a Gaussian (rather than exponential) decay.
In this way we will exhibit a simple soluble example of a 
situation where the usual Markovian \cite{Kossakowski} assumptions about 
the evolution of a quantum open system are not satisfied at any time.

Then, in Section III we will consider the more complex case with non-trivial dynamics ($\Delta\neq 0$
above). We will show that, under the same natural assumptions made in 
Section II about the distribution of coupling strengths 
in the interaction Hamiltonian, the problem can also be solved 
exactly. The solution will enable us to study two very important features
of the decoherence process. We will analyze the nature of pointer
states and also the way in which the reduced density matrix of the 
central spin evolves in time. In this case, the decay of the off--diagonal
component is not Gaussian but displays long time algebraic tails
which we obtain analytically. The most probable pointer states will be shown to 
range from eigenstates of $\sigma_z$ in the small $\Delta$ limit to 
eigenstates of $\sigma_x$ in the opposite limit of large $\Delta$ (result
that can be expected based on the considerations presented 
in \cite{pz-1999}). In Section IV we will summarize our results which, 
apart from their implications for decoherence, could 
also be relevant to quantum error correction \cite{ErrorCorrection} where 
precise knowledge of decoherence is essential to select 
an efficient strategy to defeat it.

\section{Static system - Gaussian decoherence}

Here we will consider the system described by  Eq.~(\ref{hamiltonian0}). 
We begin by outlining how to solve this model exactly, and how to find 
the time dependence of the elements of the reduced density matrix of 
the system. Let us consider an initial state for the combined 
system--environment of the form
\be
\left| \Psi_{\cal SE}(0)\right> = 
(a \left|0\right>+b \left|1\right>) \sum_{n=0}^{2^N-1} c_n \ket{n}.
\label{initialstate}
\ee
Here $\ket{n}$ are the states of the computational basis of the 
environment that diagonalizes ${\cal H_{SE}}$. The $k$-th digit of the binary form
of $n$, $n_k$, represents the state up or down in the $Z$ axis of the $k$-th spin 
of the environment. The main assumptions above are that the initial state 
is a product (no initial entanglement between the system and environment) 
and that the total state is pure. Both conditions can be easily relaxed, but  
choosing Eq.~(\ref{initialstate}) simplifies the presentation. 
The state of ${\cal SE}$ at an arbitrary time is given by:
\be
\left| \Psi_{\cal SE}(t)\right> = 
a \left|0\right> \left|{\cal E}_0 (t)\right> 
+b \left|1\right>  \left|{\cal E}_1 (t)\right>,
\label{phit}
\ee
with
\ba 
\left|{\cal E}_0 (t)\right> & = &\sum_{n=0}^{2^N-1} c_n e^{-i B_n t/2} \ket{n} \nonumber \\ 
&=& \left|{\cal E}_1 (-t)\right>,
\label{environ}
\ea
and where
\be
B_n=\sum_{k=1}^N (-1)^{n_k}g_k
\ee
The reduced density matrix of the system is then:
\ba
\rho_{\cal S} & = & {\rm Tr} _{\cal E} \left| \Psi_{\cal SE}(t)\right> 
\left< \Psi_{\cal SE}(t)\right| \nonumber \\
& = & |a|^2 \left|0\right>\left<0\right|+ a b^{*} r(t)
\left|0\right>\left<1\right| \nonumber \\ 
& + & a^{*} b r^{*}(t) \left|1\right>\left<0\right| + |b|^2 \left| 1 \right> \left< 1 \right|,
\label{reducedrho}
\ea
where the {\it decoherence factor}
$r(t)=\left<{\cal E}_1 (t)|{\cal E}_0 (t)\right>$
can be readily obtained:
\ba
r(t)&=& \sum_{n=0}^{2^N-1} \left|c_n\right|^2 e^{-i B_n t}  .
\label{roft}
\ea

It was shown in \cite{Zurek82} (using some simplifications to be discussed below) 
that for $t>0$, $r(t)$ decays rapidly
to zero, so that the typical fluctuations of the off-diagonal terms of
$\rho_{\cal S}$ will be small for large environments.  
Therefore, the decoherence factor tends to zero $\left<|r(t)|^2 \right> \underset{N\rightarrow \infty}{\longrightarrow} 0$,
leaving $\rho_{\cal S}$ approximately diagonal in a mixture of the pointer states 
$\left\{ \left|0\right>, \left|1\right> \right\}$ which retain preexisting
classical correlations.

We will show in this section that, for a fairly generic set of assumptions, 
the form of $r(t)$ can be
further evaluated and that -- quite universally -- it turns out to be 
approximately Gaussian in time. To prove this we will only require that
the couplings $g_k$ of Eq.~(\ref{hamiltonian0}) are 
sufficiently concentrated near
their average value so that their standard deviation
$\left<(g_k-\left<g_k\right>)^2\right>$ exists and is finite. When this condition
is not fulfilled other sorts of time dependence become possible. In particular,
$r(t)$ may be exponential when the distribution of couplings is, for example, Lorentzian.

To obtain our result we rewrite Eq.~(\ref{roft}) as
\be
r(t)=\int e^{-iBt} \eta(B) dB,
\label{rLDOS}
\ee
that is, the decoherence factor is the Fourier transform of a characteristic function
\be
\eta(B) = \sum_{n=0}^{2^N-1}  \left|c_n\right|^2 \delta(B-B_n).
\label{defeta}
\ee
Eq.~(\ref{defeta}) is a particular case of the more general strength function or local density of states \cite{LDOS},
\be
\eta(B)= \sum_\lambda |\braket{\phi_\lambda}{\Psi_{\cal SE}}|^2 \delta(B-B_\lambda),
\ee
where $\ket{\phi_\lambda}$ are the eigenfunctions of the full Hamiltonian with eigenenergies $B_\lambda$.

The discussion of decoherence in our model is thus directly related to 
the study of the characteristic function of the distribution of 
coupling energies $\eta(B)$.
Since the $B_n$'s are sums of $g_k$'s (that we assume independent of
each other), equation (\ref{rLDOS}) makes $r(t)$ itself a product of characteristic functions of the 
distributions of the couplings $g_k$. Thus, the distribution of $B_n$ belongs to the
class of the so--called {\it infinitely divisible distributions} \cite{Gnedenko,breiman}. 
The behavior of the decoherence factor $r(t)$ 
-- characteristic function of an infinitely divisible distribution --
depends only on the average and variance of the distributions of couplings
weighted by the initial state of the environment
\cite{Gnedenko,breiman}.

Assuming that the variance of the 
couplings $g_k$ is finite, we claim that for reasonable assumptions 
on the initial state of the environment (the coefficients $c_n$), and $N$ sufficiently large,
$\eta(B)$ has in general a Gaussian form. Therefore, the decoherence 
factor decays as a Gaussian with time. 
We will show this behavior with some examples where an exact solution is possible,
and discuss the regime of validity of the conjecture. 

\begin{figure}
\centering \leavevmode
\epsfxsize 3.2in
\epsfbox{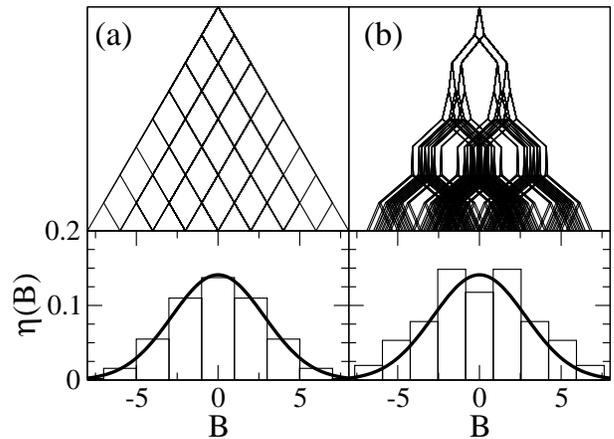}
\caption{The distribution of the energies obtains from the random walks with the
steps given by the coupling size and in the direction ($+g_k$ or $-g_k$) biased
by the probabilities $|\alpha_k|^2$ and $|\beta_k|^2$ as in Eq.~(\ref{defeta})
(although in these examples we set $|\alpha_k|^2=1/2$).
(a) When all the couplings
have the same size $g_k=g$ (Eq.~(\ref{binomial})), 
a simple Newton's triangle leads to an approximate
Gaussian for the distribution of energies. (b) When the couplings
differ from step to step (Eq.~(\ref{etageneral})), 
the resulting distribution still
has a approximately Gaussian envelope for large $N$. }
\label{Figure1} 
\end{figure}

Let us consider first the simplest case where all couplings are equal, 
$g_k=g$, and all the spins of the environment have the same initial state,
\be
\left| \Psi_{\cal SE}(0)\right> = (a \ket{0}+b \ket{1}) \bigotimes_{k=1}^N
(\alpha_k \left|0\right>_k + \beta_k \left|1\right>_k),
\label{phi0k}
\ee
with $\alpha_k=\alpha$ and $\beta_k=\beta$ for all $k$.
The decoherence factor then takes the simple form $r(t)~=~
( |\alpha|^2 e^{i g t} + |\beta|^2 e^{-i g t})^N$. Expanding this expression we obtain 
\be
r(t) = \sum_{k=0}^{N} \binom{N}{k} |\alpha|^{2k} |\beta|^{2(N-k)|}
\exp{[- i g t (2k-N)]}.
\label{binomial}
\ee
As follows from the Laplace-de Moivre theorem \cite{Gnedenko}, 
for sufficiently large $N$ the coefficients of the binomial expansion
of Eq. (\ref{binomial}) can be approximated by a Gaussian,
\be
\binom{N}{k} |\alpha|^{2k} |\beta|^{2(N-k)|} \simeq 
\frac{e^{-\frac{(k-N|\alpha|^2)^2}{2 N |\alpha \beta|^2}}}{\sqrt{2 \pi N |\alpha \beta|^2}}.
\ee
Therefore for large $N$
\be
\eta(B) \simeq \frac{\exp {\left[-\frac{\left[B/g-N(|\alpha|^2-|\beta|^2)\right]^2}{8 N |\alpha \beta|^2}\right]}}
{\sqrt{8 \pi N |\alpha \beta|^2}}.
\label{Ldosrw}
\ee
This generic behavior can be interpreted as a result of the law of large numbers \cite{Gnedenko}:
the energies $B_n$ of the composite ${\cal SE}$ system can be thought of as being the terminal points of an $N$--step random walk.
The contribution of the $k$--th spin of the environment to the random energy is $+g$ or
$-g$ with probability $|\alpha|^2$ or $|\beta|^2$ respectively [Fig.~(\ref{Figure1}.a)]. Therefore, 
the set of all the resulting energies must have an 
(approximately) Gaussian distribution. 

We can carry out the same argument in the more general case of Eq.~(\ref{phi0k}) for different couplings and initial states for the spins of the environment. Here,
\be
r(t) = \prod_{k=1}^N ( |\alpha_k|^2 e^{i g_k t} + |\beta_k|^2 e^{-i g_k t}).
\ee
The ``random walk'' picture that yielded the distribution of the couplings
remains valid [see Fig.~(\ref{Figure1}.b)].
However, now the individual steps in the random walk are no longer all equal. Rather,
they are given by the set $\left\{g_k \right\}$ [see Eq.~(\ref{hamiltonian0})]
with each step $g_k$ taken just once in a given walk.
There are $2^N$ such distinct random walks $W_n$, one for every state $\ket{n}$ of the environment. Each walk contributes to $\eta(B)$ with the weight given by 
the product of the relevant $|\alpha_k|^2$ and $|\beta_k|^2$, or 
right ($k\in W^+_n$) and left ($k\in W^-_n$)
``steps" respectively.  
The weight of the $n$-th walk is then given by
\be
|c_n|^2 = \left( \prod_{k\in W_n^+} |\alpha_k|^2 \right) 
 \left(\prod_{k\in W_n^-}|\beta_k|^2 \right).
 \label{cns}
\ee
The terminal points $B_n$ of the random walks may or may not be degenerate: 
As seen in Fig.~(\ref{Figure1}), in the degenerate case, the whole
collection of $2^N$ random walks ``collapses" into $N+1$ terminal
energies. More typically, in the non-degenerate case [also displayed in Fig.~(\ref{Figure1})],
there are $2^N$ different terminal energies $B_n$. 
In any case, the ``envelope" of the distribution $\eta(B)$ will be Gaussian, 
as we shall argue below.

\begin{figure}
\centering \leavevmode
\epsfxsize 3.2in
\epsfbox{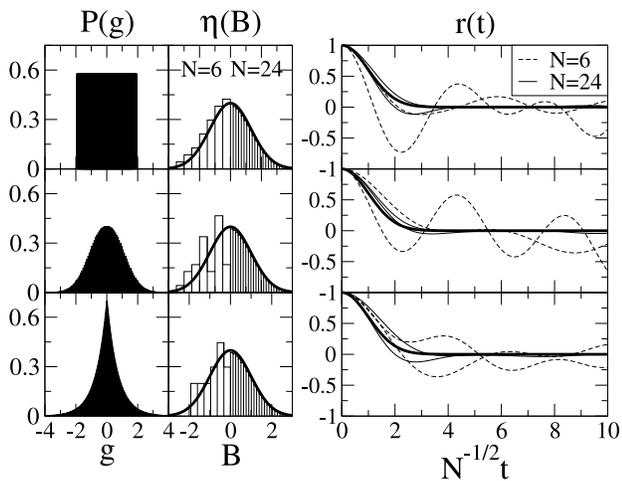}
\caption{Assumed distribution of the couplings $g_k$,
and resulting distribution of the
eigenenergies $B_n$ (left panels) for $N=6$ and $N=24$. 
In the case of $|\alpha_k|^2=1/2$ this distribution is in effect the
``strength function" (local density of states). The corresponding  
decoherence factors $r(t)$ for different initial
conditions with $N=6$ (dashed lines), $N=24$ (thin solid lines) and the average
(bold line) is shown on the right. Note the rapid convergence to a Gaussian behavior.}
\label{Figure2} 
\end{figure}

Let us compute the characteristic function $\eta(B)$. If we
denote $x_k$ the random variable that takes the value $+g_k$ or $-g_k$ with probability
$|\alpha_k|^2$ or $|\beta_k|^2$ respectively, then its mean value
$a_k$ and its variance $b_k$ are
\ba
a_k&=&(|\alpha_k|^2-|\beta_k|^2)g_k, \nonumber \\
b_k^2&=&g_k^2-a_k^2=4|\alpha_k|^2|\beta_k|^2g_k^2.
\ea
The behavior of the sums of $N$ random variables $x_k$ (and thus, of their characteristic
function) depends on whether the so--called Lindeberg condition holds
\cite{Gnedenko}. It is expressed in terms of the cumulative variances
$s_N^2=\sum b_k^2$, and it is satisfied when the probability of the large
individual steps is small; e.g.:
\be
P(\underset{1\le k \le N}{{\rm max}} |g_k-a_k| \ge \tau s_N)
\underset{N\rightarrow\infty}{\longrightarrow} 0,
\ee
for any positive constant $\tau$. In effect, Lindeberg condition demands that $s_N$ be
finite: when it is met, the resulting distribution
of energies $B=\sum x_k$ is Gaussian
\be
\eta\left( \frac{B-{\overline B}_N}{s_N}<x\right) 
\underset{N\rightarrow\infty}{\longrightarrow}
\int_{-\infty}^{x} e^{-y^2/2}dy,
\ee
where ${\overline B}_N=\sum_k a_k$. This  implies
\be
\eta(B)\simeq \frac{1}{\sqrt{2\pi s_N^2}}
\exp{\left(-\frac{(B-{\overline B}_N)^2}{2 s_N^2}\right)},
\label{etageneral}
\ee
an expression in excellent agreement with numerical results already for modest
values of $N$.
After applying the Fourier transform
of Eq. (\ref{rLDOS}), this distribution of energies yields a corresponding approximately
Gaussian time--dependence of $r(t)$ [Fig.~(\ref{Figure2})]
\be
r(t)\simeq e^{i {\overline B}_N t} e^{-s_N^2 t^2/2}.
\label{rgeneral}
\ee
Moreover, at least
for short times of interest for, say, quantum error correction, $r(t)$ is
approximately Gaussian already for relatively small values of $N$. This
conclussion holds whenever the initial distribution of the couplings has a
finite variance. Note that in particular, we did not have to assume ``randomness'' of 
the couplings $g_k$ [see e.g. Eqs.~(\ref{binomial})-(\ref{Ldosrw})].

A random initial state for the environment (not necessarily a product state) instead of Eq.~(\ref{phi0k}) gives basically the same result. In this case typically $c_n \simeq 2^{-N/2} e^{i \phi_n}$, with $\phi_n$ a random 
phase between $0$ and $2 \pi$. From Eq. (\ref{defeta}), 
\be
\eta(B)\simeq\frac{1}{2^N} \sum_{k=0}^N \binom{N}{k} \delta[B-g(2k-N)],
\ee
and, as above, the Gaussian limit for large $N$ applies.

\begin{figure}
\centering \leavevmode
\epsfxsize 3.2in
\epsfbox{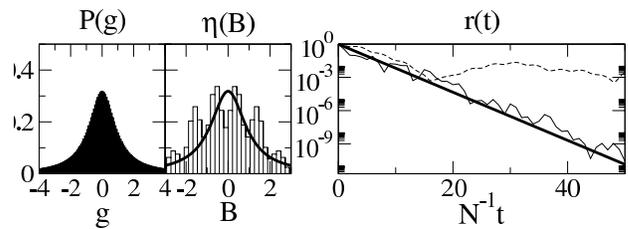}
\caption{Same as Fig. 2 but for a Lorentzian distribution of the couplings
${g_k}$. In this case $r(t)$ decays exponentially. 
The histogram and the dashed line in $r(t)$ correspond to $N=20$,
the straight thin line is a particular case for $N=100$ and the thick line is 
the average. We note that the convergence is slower than in the Gaussian case 
of Fig. 2, because realizations of ${g_k}$ are more likely to have 
one or two dominant couplings. Therefore, although the average shows a clear
exponential decay, fluctuations are noticeable even for large $N$.}
\label{Figure3} 
\end{figure}

It is also interesting to investigate cases when Lindeberg condition is not
met. Here, the possible limit distributions are given by the stable (or L\'{e}vy)
laws \cite{breiman}.
One interesting case is a Lorentzian distribution of couplings, which
yields an exponential decay of the decoherence factor [see Fig.~(\ref{Figure3})]. 
Such a distribution could be obtained for instance by considering dipolar interaction between spins randomly placed in a sample. The long range nature of the interaction gives rise to the Lorentzian distribution and therefore to the exponential decay that can be deduced by statistical arguments \cite{DiffusionNMR}.

\subsection{Relation to the Loschmidt echo}

The Fourier transform of the strength function
$\eta(B)$ is also related to the Loschmidt echo 
\cite{LETheo} (or fidelity) in the so called Fermi Golden
rule regime. The fact that the purity and the fidelity have closely related
decay rates has been recently shown \cite{LEdeco} 
for the case of a bath composed of non--interacting harmonic
oscillators. In this sense our results could be interpreted as an extension
of the discussion of Ref. \cite{LEdeco} to spin environments. 

The connection with fidelity is more easily seen if we write a
generalized version of the Hamiltonian (\ref{hamiltonian0}),
\ba
{\cal H_{SE}} =\frac{1}{2}\left(\left|0\right>\left<0\right| \otimes 
{\cal H}^0_{\cal E} + \left|1\right>\left<1\right| \otimes 
{\cal H}^1_{\cal E} \right) .
\label{hamiltonianCnot}
\ea
The decoherence factor is then the overlap of the initial state of the
environment $\left| \Psi_{\cal E}(0) \right>$ evolved with two different
Hamiltonians,
\ba
r(t)= \left< \Psi_{\cal E}(0) \right|e^{i{\cal H}^0_{\cal E} t/2} 
e^{-i{\cal H}^1_{\cal E} t/2} 
\left| \Psi_{\cal E}(0) \right>,
\ea
which clearly has the form of the amplitude of the Loschmidt echo for the
environment with the two states of the system as the perturbation. In the
 model of Eq. (\ref{hamiltonian0}), ${\cal H}_{\cal E}^0=-{\cal H}_{\cal E}^1$ and thus
\ba
r(t)&=& \left< \Psi_{\cal E}(0) \right|e^{-i{\cal H}_{\cal E}^1 t}
\left| \Psi_{\cal E}(0) \right>.
\label{autocorrelation}
\ea
This expression is the survival probability of the initial state of the
environment under the action of the Hamiltonian ${\cal H}_{\cal E}^1$, 
which is known to be the Fourier transform of the strength function
\cite{Heller}. This connection provides another way to understand
Eq.~(\ref{rLDOS}).

\section{Decoherence and dynamics}

In this section we will study the more general Hamiltonian of Eq.~(\ref{hamiltonian1}), that is we will include a self Hamiltonian to the central system. The results of the previous section will be contained in the limit of $\Delta = 0$, however we will see that  for any finite $\Delta$ the behavior of the decoherence factor will be non-trivially different from what we obtained in the previous section.

Despite its more complex appearence, the model given by Eq.~(\ref{hamiltonian1}) is still exactly solvable \cite{Dobrovitski}. Since the states $\ket{n}\bra{n}$ of the environment commute with the Hamiltonian, we can write the evolution operator for the combined system-environment as
\be
U(t)=\prod_{n=0}^{2^N-1} U_{B_n}(t) \otimes \ket{n}\bra{n},
\ee
with
\be
U_{B_n}(t)= I \cos (\Omega_n t)-i \frac{(\sigma_z B_n+\sigma_x \Delta)}{\Omega_n} \sin(\Omega_n t),
\ee
and $\Omega_n^2 = \Delta^2 + B_n^2 $.
The physical interpretation of this results is that for every state of the environment $\ket{n}$ the effective dynamics of the system is given by a magnetic field $\vec{\Omega}_n=(\Delta,0,B_n)$ in the $XZ$ plane. Seen from this perspective, the decoherence is produced by the dispersion of the fields $B_n$.

\begin{figure}[rt]
\centering \leavevmode
\epsfxsize 3.2in
\epsfbox{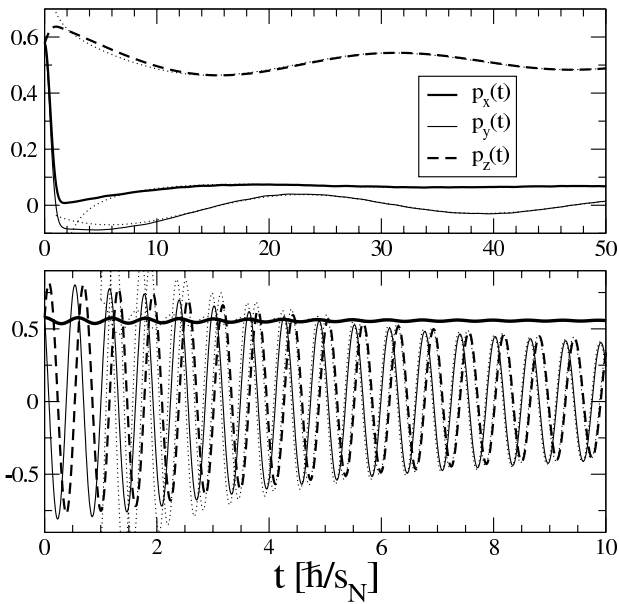}
\caption{Behavior of the components of the polarization vector for long times for $\Delta/s_N=0.1$ (top panel) and $\Delta/s_N=5$ (bottom panel). Numerical results in solid lines and analytical predictions in dashed lines.}
\label{Figure4} 
\end{figure}

The reduced density matrix of the system at an arbitrary time $t$ is
\be
\rho(t)=\sum_{n=0}^{2^N-1} |c_n|^2 U_{B_n}(t) \rho(0) U_{B_n}^\dagger(t),
\ee
or, transforming the notation and using Eq.~(\ref{defeta}),
\be
\rho(t)=\int  U_B(t) \rho(0) U_B^\dagger(t) \ \ \eta(B) \ \ dB.
\label{rhotdelta}
\ee

For simplicity, we will work with the polarization vector $\vec{p}$, such that $\rho=(I+\vec{p} \cdot \vec{\sigma})/2$. Thus, 
\be
\vec{p}(t)=\int \vec{p}(t,B) \ \eta(B) \ dB
\label{polarization}
\ee
For an arbitrary time $t$, we find
\begin{widetext}
\begin{subequations}
\be
p_x(t,B) = p_x(0) \frac{\Delta^2+B^2 \cos (2 \Omega_B t)}{\Omega_B^2} 
- p_y(0) \frac{B}{\Omega_B} \sin(2 \Omega_B t)
+ p_z(0) \frac{2 \Delta B }{\Omega_B^2}\sin^2(\Omega_B t) ,
\label{pxtotal}
\ee
\be
p_y(t,B) = p_y(0) \cos(2 \Omega_B t) + \frac{\sin(2 \Omega_B t)}{\Omega_B} \left[ p_x(0) B - \Delta p_z(0) \right],
\label{pytotal}
\ee
\be
p_z(t,B) = p_z(0) \frac{B^2+\Delta^2 \cos (2 \Omega_B t)}{\Omega_B^2} 
+ p_x(0) \frac{2 \Delta B }{\Omega_B^2}\sin^2(\Omega_B t) 
+ p_y(0) \frac{\Delta}{\Omega_B} \sin(2 \Omega_B t),
\label{pztotal}
\ee
\label{ptotal}
\end{subequations}
\end{widetext}

According to the results of the previous section, in general for large $N$ we can assume a Gaussian shape for $\eta(B)$. By using a Gaussian centered around zero,
\be
\eta(B)= \frac{1}{\sqrt{2\pi s_N^2}} \exp{\left(-B^2/2 s_N^2\right)},
\ee
Eqs. (\ref{ptotal}) simplify because the odd terms in $B$ don't contribute to the final result.

Using these assumptions, we were not able to obtain a solution of the integral in Eq.~(\ref{polarization}) for arbitrary values of $s_N$ and $\Delta$. However, we can solve the two limiting cases $s_N\gg\Delta$ and $s_N\ll\Delta$, which turn out to give non-trivial results. 

Let us consider first the case where $s_N\gg\Delta$, that is, where the central spin dynamics is so slow that its behavior should approach that obtained in the previous section. Indeed, for short times ($t\ll\Delta^{-1}$), using a Taylor expansion of Eqs. (\ref{ptotal}) around $\Delta=0$ one finds
\ba
p_x(t)&=&p_x(0) e^{-2 t^2 s_N^2} \nonumber \\
p_y(t)&=&p_y(0) e^{-2 t^2 s_N^2} -p_z(0) \frac{\Delta}{s_N} \sqrt{\frac{\pi}{2}}{\rm Erf}(\sqrt{2} s_N t) \nonumber \\
p_z(t)&=&p_z(0) + p_y(0) \frac{\Delta}{s_N} \sqrt{\frac{\pi}{2}}{\rm Erf}(\sqrt{2} s_N t) ,
\label{pdelta0smallt}
\ea
where ${\rm Erf}(x)$ is the error function. To obtain the long time behavior, we need to perform the integrals on $B$ by stationary phase approximation. In the limit $t \gg \Delta^{-1}$ we find
\begin{widetext}
\ba
p_x(t)&\simeq& p_x(0) \left[ \gamma \left( \frac{\Delta}{\sqrt{2} s_N}\right) + 
\frac{1}{ \sqrt{8 \Delta s_N^2 t^3}} \cos \left( 2 \Delta t + \frac{3 \pi}{4} \right) \right],
\nonumber \\
p_y(t)&\simeq& \sqrt{ \frac{\Delta}{2 s_N^2 t}} 
\left[ p_y(0) \cos \left(2\Delta t+\frac{\pi}{4} \right) - p_z(0) \sin \left( 2 \Delta t+\frac{\pi}{4} \right) \right],
\nonumber \\
p_z(t)&\simeq& p_z(0) \left[ 1-\gamma \left(\frac{\Delta}{ \sqrt{2} s_N}\right) +  
\sqrt{ \frac{\Delta}{2 s_N^2 t}} \cos \left( 2\Delta t+\frac{\pi}{4} \right) \right] 
+ p_y(0) \sqrt{ \frac{\Delta}{2 s_N^2 t} } \sin \left( 2\Delta t+\frac{\pi}{4} \right),
\label{pdelta0larget}
\ea
\end{widetext}
with $\gamma (x) = \sqrt{\pi} x e^{x^2} \left(1-{\rm Erf}(x) \right)$. In this limit, $\gamma \left(\frac{\Delta}{ \sqrt{2} s_N}\right) \ll 1$

Note that for any $\Delta \neq 0$ the $X$ component of the polarization does not decay to zero, indicating that the decoherence process is not completely effective in this direction [the $Y$ component does go to zero for large times due to the symmetry of Hamiltonian (\ref{hamiltonian1})]. Also, note that even a small self-Hamiltonian of the system always ends up turning a fast (Gaussian) decay into a slow (power law) one. 

In the opposite limit of strong self-dynamics of the system, $s_N \ll \Delta$, we can obtain an expression valid for all times by expanding $\Omega_B \simeq \Delta+B^2/2\Delta$. After some algebra,
\begin{widetext}
\ba
p_x(t)&=& p_x(0) \left[ \gamma \left(\frac{\Delta}{ \sqrt{2} s_N}\right) +  \frac{s_N^2}{\Delta^2} 
\frac{\cos \left(2 \Delta t + \frac{3}{2} \arctan \frac{2 s_N^2 t}{\Delta} \right)}{\left(1+\frac{4 s_N^4 t^2}{\Delta^2} \right)^{3/4}} 
 \right]
\nonumber \\
p_y(t)&=& \frac{1}{\left(1+\frac{4 s_N^4 t^2}{\Delta^2} \right)^{1/4}}
\left[ p_y(0) \cos \left(2 \Delta t + \frac{1}{2} \arctan \frac{2 s_N^2 t}{\Delta} \right) - 
p_z(0) \sin \left(2 \Delta t + \frac{1}{2} \arctan \frac{2 s_N^2 t}{\Delta} \right)
\right]
\nonumber \\
p_z(t)&=&  p_z(0) 
\left[ 1- \gamma \left(\frac{\Delta}{ \sqrt{2} s_N}\right) 
+\frac{\cos \left(2 \Delta t + \frac{1}{2} \arctan \frac{2 s_N^2 t}{\Delta} \right)}
{\left(1+\frac{4 s_N^4 t^2}{\Delta^2} \right)^{1/4}} 
 \right] + 
p_y(0) \frac{\sin \left(2 \Delta t + \frac{1}{2} \arctan \frac{2 s_N^2 t}{\Delta} \right)}
{\left(1+\frac{4 s_N^4 t^2}{\Delta^2} \right)^{1/4}} ,
\label{pB0}
\ea
\end{widetext}
In the long time limit ($t \gg \Delta/s_N^2$) these expressions are equal to Eqs.~(\ref{pdelta0larget}), only that now $\gamma \left(\frac{\Delta}{ \sqrt{2} s_N}\right) \simeq 1$. The results above for large and small $\Delta$ agree well with numerical simulations, as shown in Fig.~(\ref{Figure4}).

\subsection{Pointer basis}

\begin{figure}
\centering \leavevmode
\epsfxsize 3.2in
\epsfbox{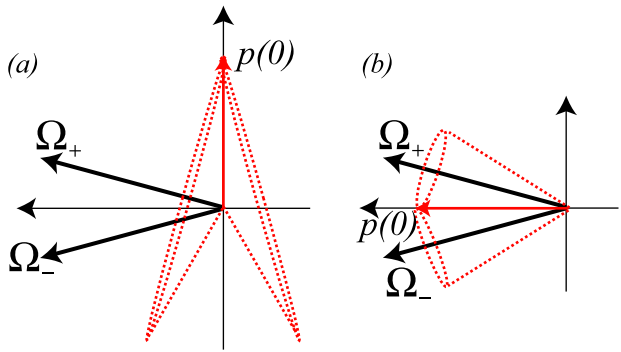}
\caption{Schematic solution of the Hamiltonian (\ref{hamiltonian1}) for two initial polarizations in the $Z$ axis (left panel) and in the $X$ axis (right panel) . Supposing that $B$ takes only two possible values, $\pm s_N$, the solution for each field $B$ is the precession of the polarization around $\vec{\Omega_{\pm}}=(\Delta,0,\pm s_N)$. The total polarization is the average of the two cones, which gives a small residue along the $Z$ axis. The polarization in $X$ is almost fully preserved.}
\label{Figure5} 
\end{figure}

The above results allow us to draw some conclusions about the nature of the decoherence process 
and the pointer states which are dynamically selected by the environment. First, we can notice that for long times the polarization vector converges 
to a certain value (which, in general, depends on $\Delta$ and other parameters of the
model). Second, we note that when $\Delta\neq 0$ the $X$ component of the polarization vector 
does not decay to zero but is resilient to the interaction with the environment. This is 
the case even if the system interacts with the environment through the $Z$ component of the spin. 

The states which are dynamically selected by the environment are dramatically different
in the two oposite regimes we examined above. For small values of $\Delta$, the eigenstates
of the $Z$ component of the central spin are pointer states. They are minimally perturbed by 
the interaction with the environment (in the previous section, where $\Delta =0$ was assumed, 
this emerged as an exact result since $p_z$ is conserved). 
However, for large $\Delta$ (i.e. $\Delta\gg s_N$), the 
fact that $p_x(t\rightarrow\infty) \simeq 1$ is a signature of the decoherence process selecting 
a completely different set of pointer states. In fact, in this case, the pointer
states turn out to be eigenstates of the system Hamiltonian, which is proportional to 
$\sigma_x$. Thus, this model enables us to examine these two very different situations:
one where the interaction with the environment dominates ($\Delta\ll s_N$) and $\sigma_z$ 
eignestates are selected; the other where the self--Hamiltonian of the system dominates
($\Delta\gg s_N$) and $\sigma_x$ eigenstates are selected. 

The regime where the Hamiltonian of the system dominates (or, more precisely, where the 
environment is much slower than the system) was analyzed in a more general contex
before \cite{pz-1999} and has a natural interpretation here: This behavior is simply 
the one corresponding to the strong decoupling regime observed in Nuclear Magnetic Resonance 
\cite{Schlichter}. There, the presence of a strong magnetic field in the $X$ or $Y$ axes 
effectively decouples the spectrally resolvable spins of a sample (whose interaction is $Z$ dominant). The 
standard picture of this decoupling regime is that by rotating the polarization rapidly enough 
around $X$, any interaction in another axis is strongly suppressed and the spins 
effectively ``decouple''.

There is an instructive physical picture to understand these results. Instead of using a 
continuos distribution for $B$, let us suppose that $B$ can only take two values, $\eta(B)=[\delta(B-s_N)+\delta(B+s_N)]/2$, with $s_N \ll \Delta$. 
The classical solution for the evolution of the polarization vector is the 
precession of $\vec{p}$ around $\vec{\Omega}_{\pm}=(\Delta,0,\pm s_N)$, as shown in 
Fig.~(\ref{Figure5}) for two possible initial conditions of $\vec{p}$. 
The polarization vector of the reduced system is the average of the two cones 
corresponding to the precession around $\vec{\Omega}_{+}$ and $\vec{\Omega}_{-}$
The presence of a small component $\Delta$ in the $Z$ axis tilts the precession 
cones so that their average is almost $1$ in the $X$ direction and has a small residue on the $Z$ axis 
(the $Y$ component cancels due to the symmetry).

\begin{figure}
\centering \leavevmode
\epsfxsize 3.2in
\epsfbox{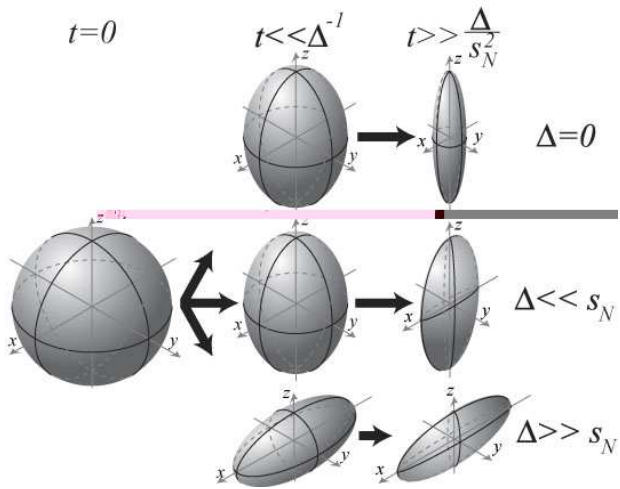}
   \caption{Bloch sphere representation of our main results. Possible initial states at $t=0$ are represented on the left by a full sphere. Intermediate times (center) are similar for $\Delta=0$ and $\Delta\ll s_N$, where decoherence reduces the Bloch sphere equally fast (Gaussian) in the $X$ and $Y$ axes. For large $\Delta$, $Y$ and $Z$ axes are decohered in a slower algebraic way. For long times, right panel, the $\Delta=0$ case is completely decohered to the $Z$ axis, while the small $\Delta$ retains some polarization along the $X$ axis. The large $\Delta$ case is almost the opposite, retaining almost all polarization in the $X$ axis and a small residue in the $Z$ axis.}
   \label{Figure6}
\end{figure}

\section{Conclusions}

We have studied a very simple model of decoherence due to spin environments. We showed that  the decoherence factor will generically have a Gaussian decay when there is no self--Hamiltonian for the system. We note that  similar behavior 
was observed for short times by studying the decoherence process in models where the largest energy scale is the system-environment interaction strength \cite{Haake}. 
A model similar to (\ref{hamiltonian0}) is used in the NMR setting \cite{DiffusionNMR} to compute corrections to the second and fourth moments of the decay of the polarization signal. Here the idea is to treat the interaction with surrounding spins as an effective local magnetic field that shifts the Larmor frequency inhomogenuosly across the sample. The statistical treatment used in \cite{DiffusionNMR} contrasts with the exact solution presented in this work, even in the presence of a self--Hamiltonian of the central spin. 
Thus, our model has applicability and relevance to a larger class of physical situations.
There is a substantial body of work 
\cite{Dobrovitski,deRaedt,Loss,Maximilian} on decoherence due to spin environments,
stimulated in part by the interests of quantum computation. Our results are most relevant for quantum error correction and other strategies to fight decoherence in a quantum computer. For example Gaussian time dependence of the decoherence factor would suggest a different (more frequent) error correction than the exponential dependence often assumed with little or no justification.

We also showed how by adding a self-Hamiltonian for the system one can dramatically change the main features of the decoherence process. Even for the case of slow dynamics 
of the system, we found that for long times the initial Gaussian behavior changes to a 
power law. On the other end, when the self-Hamiltonian is much stronger than the 
interaction with the environment, the whole process changes its nature. The decay 
is predominantly a power law. Moreover, the pointer states now correspond to 
eigenstates of the system rather than eigenstates of the system-environment 
interaction \cite{pz-1999}. For illustrative purposes, our results are summarized
schematically in Fig.(\ref{Figure6}) using the Bloch sphere representation.

Our results, though interesting, arise from a very simplified model. A logical step 
for future research is the inclusion of intra-bath interactions. The entanglement 
thus created between spin baths will surely have an impact on the amount of decoherence 
in the system \cite{Milburn}.

Possible experimental applications of our considerations are in nuclear
magnetic resonance, and in any other situation where two-level systems
interact with spin environments.
Another area of impact of our results is in the characterization of the process that 
leads to redundancy in the environment of the classical information about the 
system \cite{redundancy}. The relation between the decoherence factor and the strength 
function might prove useful in the physical setting of strongly interacting fermions, 
where it has been shown that the strength function takes a Gaussian shape \cite{Kota}.
It is our hope that
the simple analytic model described here will assist in gaining further insights
into these fascinating problems.

We acknowledge fruitful discussions with D.A.R. Dalvit, V.V. Dobrovitski, R. 
Blume-Kohout and G. Raggio. We also acknowledge partial support from  NSA grant. JPP 
received also partial support from a grant by Fundaci\'on Antorchas.

\end{document}